\documentclass{emulateapj}

\begin{document}

\shortauthors{Luhman, Stauffer, \& Mamajek}
\shorttitle{Age of AB Dor}

\title{The Age of AB Dor\altaffilmark{1}}

\author{K. L. Luhman\altaffilmark{2}, John R. Stauffer\altaffilmark{3}, and 
E. E. Mamajek\altaffilmark{2}}

\altaffiltext{1}{This publication makes use of data products
from the Two Micron All Sky Survey, which is a joint project
of the University of Massachusetts and the Infrared Processing
and Analysis Center/California Institute of Technology,
funded by the National Aeronautics and Space Administration
and the National Science Foundation.}

\altaffiltext{2}{Harvard-Smithsonian Center for Astrophysics, 60 Garden St.,
Cambridge, MA 02138, USA; kluhman, emamajek@cfa.harvard.edu.}

\altaffiltext{3}{Spitzer Science Center, Caltech MS 314-6, Pasadena, CA 91125,
USA; stauffer@ipac.caltech.edu.}

\begin{abstract}

We have derived a new age estimate for the nearby young star AB~Dor 
and have investigated the resulting implications for 
testing theoretical evolutionary models with the data reported by Close and
coworkers for the low-mass companion AB~Dor~C.
Using color-magnitude diagrams, we find that the AB~Dor moving group
is roughly coeval with the Pleiades ($\tau=100$-125~Myr) and is clearly 
older than IC~2391 ($\tau=35$-50~Myr).
In fact, based on a comparison of the kinematics of AB~Dor and the Pleiades,
we suggest that the stars identified by Zuckerman and coworkers as members 
of a moving group with AB~Dor are remnants of the large scale star-formation 
event that formed the Pleiades.
Using the age of $\tau=50^{+50}_{-20}$~Myr adopted by Close, the 
luminosity predicted by the models of Chabrier and Baraffe for AB~Dor~C
is larger than the value reported by Close, but is still within 
the quoted uncertainties.  Meanwhile, the agreement is good when
our age estimate for AB~Dor~C is adopted.
Thus, we find no evidence in the data presented by Close for AB~Dor~C to 
suggest that previous studies using the models of Chabrier and Baraffe 
and bolometric luminosity as the mass indicator have significantly 
underestimated the masses of young low-mass stars and brown dwarfs.

\end{abstract}


\section{Introduction}
\label{sec:intro}

Surveys of young open clusters and star-forming regions have identified
hundreds of very faint members that appear to be brown dwarfs \citep{bas00}.
However, the mass estimates for these objects are dependent on the validity of 
theoretical evolutionary models \citep[e.g.,][]{dm97,bar98,cha00}. 
\citet{clo05} (hereafter C05) recently tested the accuracy of the masses derived
from these models with observations of a low-mass companion to the young
nearby star AB~Dor. By combining their adaptive optics images of AB~Dor~C
and the astrometry of the primary from \citet{gui97}, they measured a mass 
of $M=0.09\pm0.005$~$M_\odot$. C05 found that 
the theoretical evolutionary models of \citet{cha00} overestimated the 
near-infrared (IR) fluxes of AB~Dor~C by roughly one magnitude. 
They concluded that the use of these models to interpret photometry
of young low-mass objects leads to masses that are 
underestimated by a factor of two and that many of the objects 
previously identified as brown dwarfs in star-forming regions and open 
clusters are instead low-mass stars.
The age for AB~Dor is important in this context because the
comparison of the observed luminosity of AB~Dor~C to theoretical
models requires an assumed age. For instance, if AB~Dor had an age
of 100~Myr instead of 50~Myr as assumed by C05,
much of the discrepancy between model and observations would disappear.
Therefore, in this paper we reexamine the age of AB~Dor.

\section{Analysis}

In adopting an age for the AB~Dor system, C05 considered the
apparent displacement of AB~Dor above the main sequence, AB~Dor's rapid rotation
and high lithium abundance, and, in particular, its membership in a moving 
group \citep{zuc04}. The age for this moving group was derived by \citet{zuc04}
from a comparison of its H$\alpha$\ emission strengths to those of 
the Tucana group and from a comparison of its three M-type members 
to isochrones for 10~Myr and the zero age main sequence (ZAMS) 
in a diagram of M$_V$ vs. $V-K_s$, which indicated
that the AB~Dor group is older than Tucana ($\tau>30$~Myr)
and younger than the ZAMS for M stars. Based on this analysis,
\citet{zuc04} reported an age of 50~Myr for the AB~Dor group.
C05 subsequently adopted this age and assigned an uncertainty
of $^{+50}_{-20}$~Myr.

In the following analysis, we reexamine the isochronal age of the AB~Dor moving 
group by including the M-type companion AB~Dor~B (Rossiter~137B), which was 
not considered by \citet{zuc04} and C05, and by comparing 
the AB~Dor group to the Pleiades and IC~2391 open clusters. 
We then use modern astrometric data to investigate the previously proposed
notion that AB~Dor shares a common origin with the Pleiades Supercluster
\citep[e.g.,][]{innis86}.

\subsection{The Isochronal Age of the AB~Dor Moving Group}
\label{sec:iso}

The evidence for a physical association between AB~Dor~A and B is 
quite compelling. First, the trigonometric parallax of AB~Dor from the 
Very Long Baseline Interferometer places it within 25~pc of the Sun, and the
spectroscopic parallax for AB~Dor~B suggests it is also within
about 25~pc of the Sun if it has an age $\gtrsim50$~Myr. 
Based on the number density of stars in the solar
neighborhood \citep{rei02a}, the probability of finding
two stars within $10\arcsec$ of each other and closer to
the Sun than 25 pc that are not physically associated with each other is 
$\sim4\times10^{-6}$. Second, the proper motion of AB~Dor is fairly large 
($\sim0\farcs1$ per year) but the separation and position angle between
AB~Dor~A and B have remained sensibly constant for 80 years.
The radial velocities of the two stars also agree to within the
measurement errors of a few km~s$^{-1}$ \citep{innis86}.
Third, both stars are demonstrably very young based on their spectroscopic,
radio, and X-ray characteristics \citep{lim93}. Taken together, the likelihood
that these two stars are unrelated is extremely small.
Therefore, nominally one should be able to obtain a better isochronal
age estimate for the AB~Dor system from AB~Dor~B than from AB~Dor 
itself, because the displacement above the ZAMS is larger at lower
masses for a given pre-main-sequence age.

The most direct, least model-dependent way to infer an isochronal
age for the members of the AB~Dor moving group is via a comparison 
to empirical isochrones defined by well-observed open clusters.
For this comparison, we select the Pleiades and IC~2391, which have
ages of 125 and 50~Myr, respectively, according to analysis of their
Li depletion boundaries \citep{stauffer98a,bar99,bar04}.
Somewhat younger ages of 100 and 35~Myr have been derived from their
upper main sequence turnoffs \citep{meynet93,mer81}.
We compiled a list of members of the Pleiades from
\citet{stauffer87} and \citet{stauffer98b} and a list of members of IC~2391
from \citet{stauffer89b}, \citet{stauffer97}, and \citet{bar04}.
For these members, we adopt the $V$ measurements from those studies and
data at $K_s$ from 2MASS. 
The well-documented error in the Hipparcos distance to the Pleiades
indicates that for clusters beyond 100~pc, accurate main-sequence distances
are preferred over Hipparcos measurements
\citep{pin98,pin04,soderblom05}. 
Thus, we adopt distances of 133 and 154~pc 
and extinctions of $A_V=0.12$ and 0.03 for the Pleiades and IC~2391, 
respectively \citep{soderblom05,for01}.

For the AB~Dor moving group, we consider the members identified by 
\citet{zuc04} as well as the components of the AB~Dor multiple system. 
We adopt the distances for these stars from \citet{per97}, except for the two
stars that lack Hipparcos measurements, which are excluded.
We use the Johnson $V$ data compiled by \citet{per97} for all stars except
AB~Dor~A and B, for which we take $V$ from \citet{cameron97}.
Measurements at $K_s$ are adopted from 2MASS for all stars. 
Using the internal 2MASS database, D. Kirkpatrick kindly checked the accuracy 
of the 2MASS data for AB~Dor~A and B, which are separated by $9\arcsec$.
He found that the two stars are well-resolved in the short 
``Read 1" exposures and that the photometric measurements from two separate
observations agree within a few percent for both stars. 
AB~Dor~B itself is a close binary (C05).
We have estimated the individual $V$ and $K_s$ magnitudes of the components 
by combining the measured flux ratio at $K_s$ (C05), the unresolved 
photometry, an assumption of coevality, and the relation between $\Delta V$
and $\Delta K_s$ implied by the empirical isochrone in the form of the
Pleiades sequence. The uncertainties in the distance, combined photometry,
and flux ratio for AB~Dor~Ba and Bb produce one sigma errors of 
$\pm0.22$ and $\pm0.1$~mag in their individual values of $V-K_s$ and $K_s$, 
respectively.

\begin{figure}
\epsscale{0.63}
\plotone{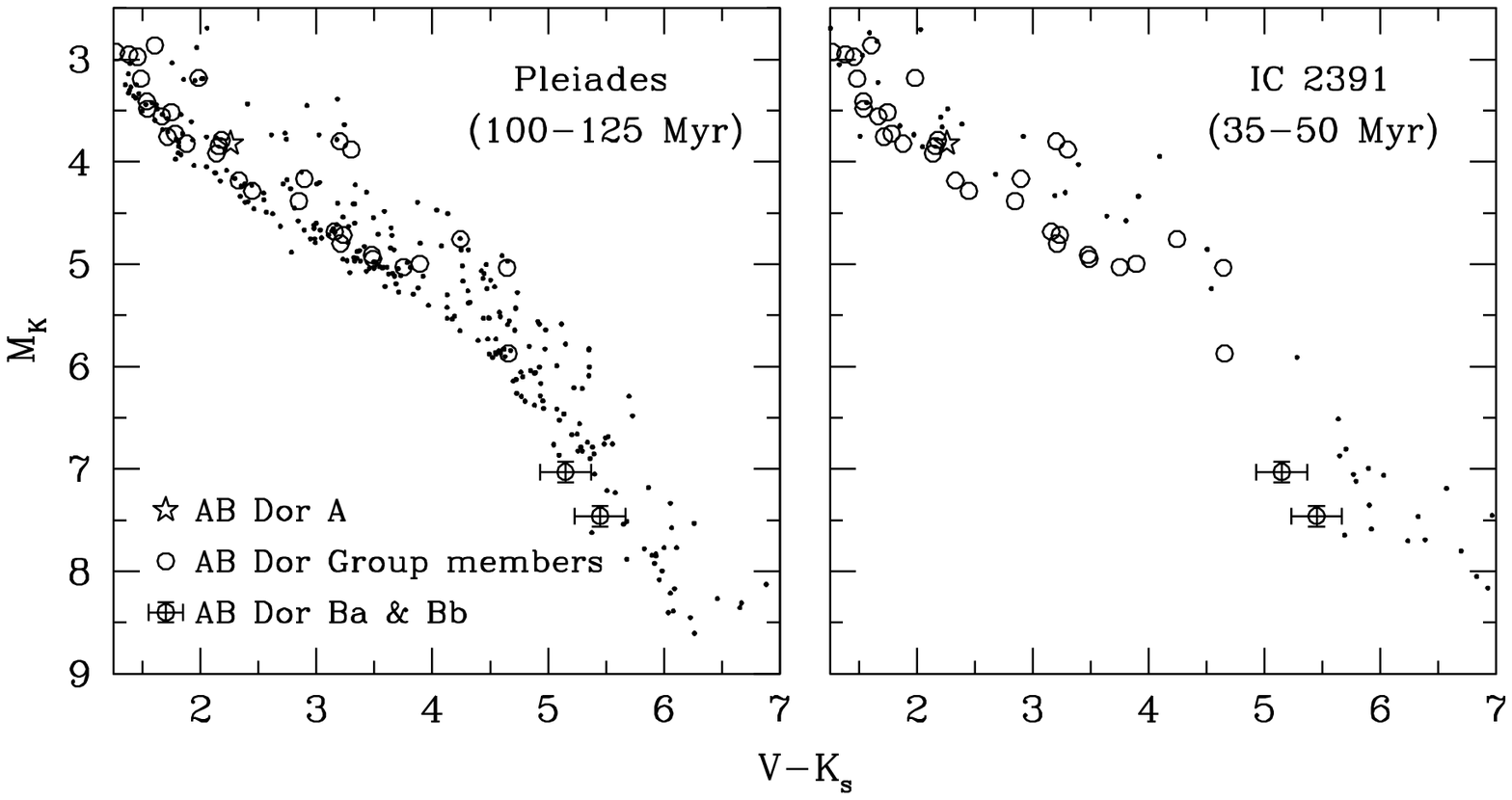}
\caption{
M$_K$ versus $V-K_s$ for the nearby star AB~Dor ({\it star}), its companions 
Ba and Bb ({\it circles with error bars}), and the members of the AB~Dor 
moving group \citep[{\it circles};][]{zuc04}.
The sequence formed by these stars coincides with the sequence for the 
Pleiades \citep[{\it left}, $\tau=100$-125~Myr,][]{meynet93,stauffer98a}
and falls well below the sequence for IC~2391
\citep[{\it right}, $\tau=35$-50~Myr,][]{mer81,bar99,bar04}.
The vertical dispersions of the open cluster sequences are
primarily due to the presence of binaries, with observational
error being a very minor contributor.
\label{fig:vk}
}
\end{figure}

\begin{figure}
\epsscale{0.6}
\plotone{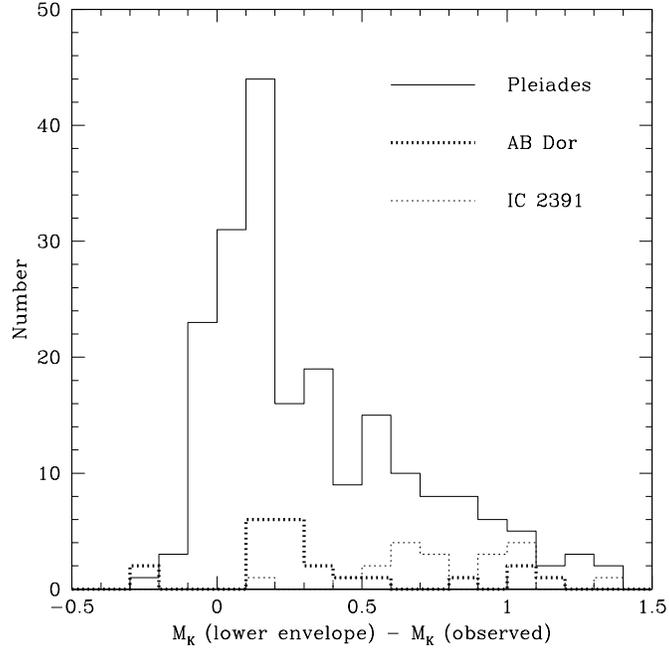}
\caption{
Histograms of $M_K$ offsets in Figure~\ref{fig:vk} 
between the lower envelope of the Pleiades sequence
and the observed positions of stars in the Pleiades 
({\it solid}), the AB~Dor moving group ({\it thick dotted}), and IC~2391
({\it thin dotted}).
The mean offsets are $<\Delta M_K>=0.34$, 0.35, and 0.83~mag, respectively.
\label{fig:dm}
}
\end{figure}

The data for the Pleiades and IC~2391 open clusters and the AB~Dor moving 
group are plotted together in a diagram of M$_K$ versus $V-K_s$ in 
Figure~\ref{fig:vk}. We find that the K and M-type members of the
AB~Dor group on average fall well below the sequence of IC~2391. 
These data clearly demonstrate that the AB~Dor moving group is older 
than IC~2391 ($\tau=35$-50~Myr). 
Meanwhile, the sequence for the AB~Dor group closely matches that of the 
Pleiades. To quantitatively compare these sequences, we have 
measured the offset in $M_K$ between the observed position of each star
and a fit to the lower envelope of the Pleiades sequence
and have generated a histogram of these offsets for each population. 
We consider only stars at $V-K_s=2$-5.5 because bluer stars have very small
displacements above the ZAMS for the ages in question and redder stars are
not present in the known membership of the AB~Dor group. 
A visual comparison of these histograms in Figure~\ref{fig:dm} indicates 
an offset of 0-0.1~mag between the AB~Dor group and the Pleiades, which
suggests that AB~Dor is coeval with the Pleiades or slightly younger. 
An offset of 0.1~mag for the AB~Dor group would correspond to an age of 
90-100~Myr if we use an age of 100-125~Myr for the Pleiades and 
the differential luminosities predicted by \citet{bar98}. 
However, the mean offsets of the Pleiades and AB~Dor sequences are 
indistinguishable, with $<\Delta M_K>=0.34$ and 0.35~mag, respectively.
In addition, AB~Dor~Ba and Bb are clearly not younger than the Pleiades,
and these stars should be given the greatest weighting in this exercise
given that they are much more directly and unambiguously associated with AB~Dor
than the moving group members and that they are better age indicators on
a color-magnitude diagram than earlier type stars. The use of
photometry in other bands and colors, such as $V$ and $V-I$, produces the
same position of AB~Dor~Ba and Bb relative to the Pleiades. 
Note that C05 reported that AB~Dor~C is overluminous relative to the Pleiades; 
this apparent discrepancy is resolved by a forthcoming analysis of the 
spectral classification of AB~Dor~C (K. L. Luhman et al., in preparation).

As for AB~Dor itself, previous studies appear to rule out binarity as an 
explanation for
its overluminous nature relative to the Pleiades, but its extremely rapid 
rotation is a plausible cause. Such effects on the H-R diagram have
been predicted from evolutionary models of solar-type stars \citep{sil00},
although they have not been obvious in observations to date \citep{stauffer87}.
We also note that at least one solar-type member of the Pleiades, HZ~102, 
is like AB~Dor~A in that its overluminous position on the color-magnitude
diagram appears not to be due to binarity. Thus, strictly from an empirical
point of view, AB~Dor appears to be consistent with the Pleiades on a
color-magnitude diagram.

\begin{figure}
\epsscale{0.6}
\plotone{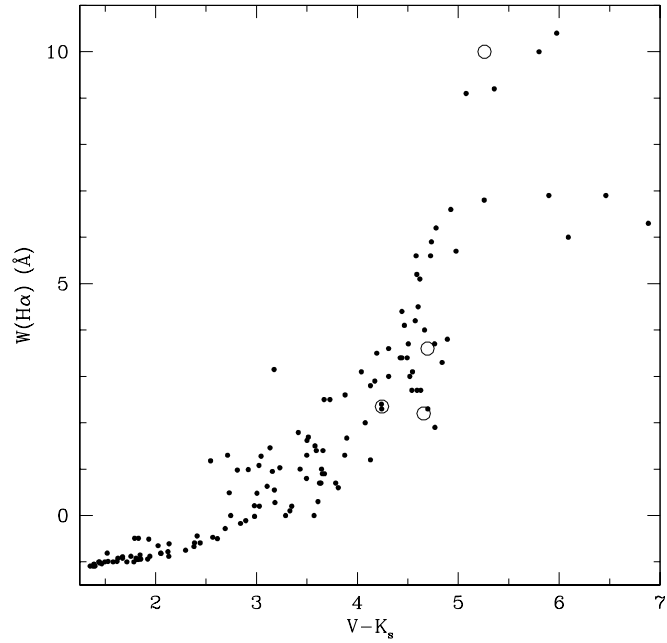}
\caption{
Equivalent widths of H$\alpha$ versus $V-K_s$
for AB~Dor~B ({\it upper circle}), the three additional 
M-type members of the AB~Dor moving group ({\it lower circles}), and 
members of the Pleiades ({\it points}).
Positive equivalent widths represent emission.
}
\label{fig:ha}
\end{figure}

The Li equivalent width and v~sin~i of AB~Dor are similar to the
values observed for rapidly rotating K dwarfs in the Pleiades \citep{sod93}
\citep[as well as in IC~2391,][]{stauffer89b,ran01}.
In addition, the H$\alpha$ emission strengths for the M-type members of the
AB~Dor group \citep{vil91,zuc04} are indistinguishable from measurements
in the Pleiades \citep{sod93,jon96,ter00,opp97,stauffer87},
as illustrated in Figure~\ref{fig:ha}
Thus, these age diagnostics are consistent with the coevality of the
AB~Dor group and the Pleiades that is suggested by the color-magnitude diagram.
Based on the above considerations, a conservative age range for
the AB~Dor multiple system and moving group is 75-150~Myr.

\subsection{The Kinematic Origin of the AB Dor Moving Group}

To calculate the Galactic space motion of AB~Dor, we use the equations
from \citet{Johnson87}, the weighted mean radial velocity from studies which 
measured the quantity multiple times 
\citep[+$28.5\pm0.6$,][]{Cameron82,Vilhu87,Balona87,Innis88,Donati97,Nordstrom04},
the long-baseline proper motion from the Tycho-2 catalog \citep{Hog00}, and 
the weighted mean of the Hipparcos and Tycho-1 trigonometric parallaxes 
\citep{per97}. We derive a heliocentric Galactic space motion vector for AB~Dor 
of ($U, V, W$) = (-7.7, -26.0, -13.6)\,$\pm$\, (0.4, 0.4, 0.3)~km~s$^{-1}$. 
For comparison, we calculated a velocity vector for the
Pleiades using the new distance estimate from \citet[][]{soderblom05} and the 
proper motion, radial velocity, and mean cluster position from 
\citet{Robichon99}, arriving at
($U, V, W$) = (-6.6, -27.6, -14.5)\,$\pm$\,(0.4, 0.3, 0.3)~km~s$^{-1}$. 
An immediate result of this analysis is that 
AB~Dor is only $\sim$2 km~s$^{-1}$ from the Pleiades in velocity space. 
But how close is AB~Dor to the Pleiades space motion compared to stars in 
the field? For a field sample, we use the 13,222 stars with calculated UVW 
velocity vectors from the magnitude-limited sample of F/G-type stars from 
\citet{Nordstrom04}.
AB~Dor is the fourth closest star to the Pleiades in terms of velocity using
the vector from \citet{Nordstrom04} (it is present in the sample 
despite its K spectral type),
while it is the eighth closest star according to our revised vector.
Many of the AB~Dor group stars from \citet{zuc04} are in the sample from
\citet{Nordstrom04}, and are among the $\sim40$ closest stars to the Pleiades 
mean motion. These stars include HD~45270, HD~19183, PW~And, and UY~Pic.
Out of the 40 stars closest to the Pleiades vector from the Nordstrom catalog,
13 are listed as members of the AB~Dor moving group by \citet{zuc04}.
Hence, AB~Dor and its moving group members are in rather exclusive company 
among the $\sim$0.3\% of stars in the Nordstrom catalog that are nearest
the Pleiades velocity vector. 

Kinematic analysis of the Hipparcos catalog \citep{per97} has
shown that many young BAF-type stars have motions clustered near 
that of the Pleiades cluster \citep[e.g.][]{Chereul98,Chereul99,Asiain99}, 
which has been referred to as the Local Association or the Pleiades
Supercluster \citep{Eggen70,Eggen75}.
\citet{innis86} suggested that AB~Dor might belong to the Pleiades 
Supercluster, while more recently \citet{zuc04} has proposed the existence 
of a moving group with AB~Dor.
\citet{zuc04} reasoned that the more negative V ($\sim-27$~km~s$^{-1}$) 
and W ($\sim-14$~km~s$^{-1}$) velocity component of the AB~Dor group makes it
distinctive from other recently discovered young stellar groups within 100~pc
of the Sun, but AB~Dor's velocity is directly coincident with the Pleiades, 
as we have shown. Very young open clusters are often found
to be surrounded by unbound OB associations \citep{Garmany94}, a situation
observed today with the $\alpha$ Persei cluster \citep{deZeeuw99}. 
We suggest that the systems identified by \citet{zuc04} as members of a moving
group with AB~Dor are probably the remnants of unbound OB and T associations
associated with the star-formation event that formed the bound Pleiades open
cluster.

\section{Discussion}

We now use our new estimate of the age of AB~Dor in
testing the theoretical evolutionary models of \citet{cha00}.
C05 tested these models by comparing the observed and predicted
values of $M_J$, $M_H$, $M_K$, and $T_{\rm eff}$ for AB~Dor~C.
However, the predicted near-IR magnitudes are subject to known deficiencies 
in the opacities used in the synthetic spectra. Indeed, discrepancies between 
observed and synthetic colors and magnitudes at near-IR wavelengths have been 
noted in previous studies \citep[e.g.,][]{cha00,leg01}.  
Instead, the most fundamental and robust parameter predicted by any set 
of models is the bolometric luminosity.
In addition, the masses of young low-mass objects are typically inferred
from the predicted luminosities rather than near-IR magnitudes.
Therefore, we consider this parameter in the following discussion.

\begin{figure}
\epsscale{0.6}
\plotone{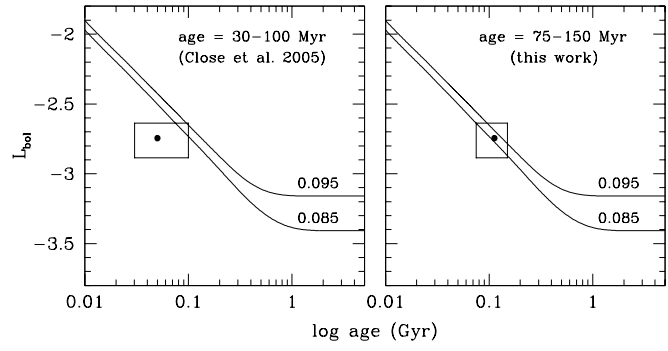}
\caption{
Comparison of the luminosity of AB~Dor~C reported by C05 to the 
values predicted by the evolutionary models of \citet{cha00} for masses
bracketing its dynamical mass of $0.09\pm0.005$~$M_\odot$ (C05).
{\it Left:} Using the age adopted by C05,
the models appear to overestimate the luminosity of AB~Dor~C, although the
observed and predicted values overlap within the measurement uncertainties
({\it rectangles}). {\it Right:} The agreement is better with our age estimate.
}
\label{fig:lbol}
\end{figure}

To test the predicted luminosity for AB~Dor~C, we use the diagrams of 
luminosity versus age in Figure~\ref{fig:lbol}. We show the luminosity
of $0.0018\pm0.0005$~$L_\odot$ derived by C05 with 
the estimates of age for AB~Dor C from C05 and from this work.
Although C05 reported a significant difference between the
observed and predicted near-IR fluxes for AB~Dor~C (primarily $J$ and $H$), 
Figure~\ref{fig:lbol} demonstrates that the predicted bolometric luminosity 
is actually consistent with their measured value within the uncertainties.
Meanwhile, using our estimate of the age of AB~Dor~C, 
the data and model predictions agree reasonably well. 
The claim by C05 that the masses of young low mass stars and brown
dwarfs are significantly underestimated using the models of \citet{cha00} 
hinged on the assumed age for AB~Dor~C (as well as its other stellar 
properties). Because our best estimate for the age of the AB Dor system is 
twice as old as they assumed, we find no significant discrepancy with the
model predictions when using bolometric luminosity as the mass indicator.

\acknowledgments
We are grateful to Davy Kirkpatrick for examining the 2MASS photometry
for AB~Dor~B.  We also thank Sandy Leggett and Laird Close for their 
comments on the manuscript.
K. L. was supported by grant NAG5-11627 from the NASA Long-Term Space 
Astrophysics program. E. M. was supported by a Clay Postdoctoral Fellowship 
from the Smithsonian Astrophysical Observatory.

\end{document}